\documentclass[conference]{IEEEtran}
\IEEEoverridecommandlockouts
\usepackage{cite}
\usepackage{amsmath,amssymb,amsfonts}
\usepackage{algorithmic}
\usepackage{graphicx}
\usepackage{textcomp}
\usepackage{xcolor}

\usepackage{enumitem}
\setlist[itemize]{first=\small}

\usepackage{acronym}
\newacro{AOP}[AOP]{Aspect-oriented Programming}
\newacro{COP}[COP]{Context-oriented Programming}
\newacro{HPC}[HPC]{High-Performance Computing}
\newacro{AST}[AST]{abstract syntax tree}
\newacro{JIT}[JIT]{Just-In-Time}
\newacro{JPM}[JPM]{JoinPoint Model}
\newacro{JP}[JP]{JoinPoint}
\usepackage[compatibility=false]{caption}
\usepackage[newfloat,frozencache=true,cachedir=_minted]{minted}
\setminted{fontsize=\scriptsize}
\setmintedinline{fontsize=\small}

\DeclareFloatingEnvironment[placement=ht!, name={Code}]{code}

\usepackage[many]{tcolorbox}
\newtcolorbox{mytcolorbox}{arc=0pt,boxsep=0pt,left=2mm,right=2mm,top=2mm,bottom=2mm}
\BeforeBeginEnvironment{minted}{\begin{mytcolorbox}}
\AfterEndEnvironment{minted}{\end{mytcolorbox}}

\def\BibTeX{{\rm B\kern-.05em{\sc i\kern-.025em b}\kern-.08em
    T\kern-.1667em\lower.7ex\hbox{E}\kern-.125emX}}
\begin{document}

\title{Aspect-oriented Programming with Julia}

\author{\IEEEauthorblockN{1\textsuperscript{st} Osamu Ishimura}
  \IEEEauthorblockA{
    \textit{The University of Tokyo}\\
    Tokyo, Japan \\
    oishimura@is.s.u-tokyo.ac.jp}
  \and
  \IEEEauthorblockN{2\textsuperscript{nd} Yoshihide Yoshimoto}
  \IEEEauthorblockA{
    \textit{The University of Tokyo}\\
    Tokyo, Japan \\
    yosimoto@is.s.u-tokyo.ac.jp}
}

\makeatletter
\def\ps@IEEEtitlepagestyle{
  \def\@oddfoot{\mycopyrightnotice}
  \def\@evenfoot{}
}
\def\mycopyrightnotice{
  {\footnotesize
      \begin{minipage}{\textwidth}
        \centering
        \textcopyright 2024 IEEE. Personal use of this material is permitted. Permission from IEEE must be obtained for all other uses, in any current or future media, including reprinting/republishing this material for advertising or promotional purposes, creating new collective works, for resale or redistribution to servers or lists, or reuse of any copyrighted component of this work in other works.
      \end{minipage}
    }
}

\maketitle

\begin{abstract}
  This paper proposes integrating \ac{AOP} into Julia, a language widely used in scientific and \ac{HPC}.
  \ac{AOP} enhances software modularity by encapsulating cross-cutting concerns, such as logging, caching, and parallelizing, into separate, reusable \textit{aspects}.
  Leveraging Julia's powerful metaprogramming and \ac{AST} manipulation capabilities, we introduce AspectJulia, an \ac{AOP} framework designed to operate within Julia's runtime environment as a package.

  AspectJulia enables developers to define and apply \textit{aspects} seamlessly, leading to more modular, maintainable, and adaptable code.
  We detail the implementation of AspectJulia and present diverse use cases, ranging from \ac{HPC} and scientific computing to business applications, demonstrating its effectiveness in managing cross-cutting concerns.
  This integration simplifies application development and improves the adaptability of existing Julia modules and packages, paving the way for more efficient and maintainable software systems.
\end{abstract}

\begin{IEEEkeywords}
  Aspect-oriented Programming, Julia, AspectJulia, Programming Language
\end{IEEEkeywords}

\section{Introduction}
\ac{AOP} is a programming paradigm that enhances software modularity to simplify development and improve maintainability.
Although \ac{AOP} has been successfully integrated into web development frameworks like Spring AOP~\cite{AspectOr56:online}, its adoption in HPC still needs to be improved.
While AspectJ~\cite{10.1007/3-540-45337-7_18} and AspectC++~\cite{10.5555/564092.564100} have been used in some HPC applications, they have yet to be widely adopted.
One significant factor is the need for robust \ac{AOP} implementations in languages commonly used for \ac{HPC} applications.
This study introduces integrating \ac{AOP} into Julia, a programming language extensively used in \ac{HPC} applications and other fields.
Julia is favored for its performance on par with compiled languages, even though it is dynamic.
This paper discusses the benefits of incorporating \ac{AOP} into Julia and the technical aspects of its implementation. Then it illustrates the practicality and effectiveness of \ac{AOP} in Julia through diverse use cases, including both \ac{HPC} and business applications.

\section{Background}

\subsection{What is \ac{AOP}}
\ac{AOP} is a programming paradigm proposed by Gregor Kiczales et al.~\cite{10.1007/BFb0053381} to enhance modularity by separating and encapsulating cross-cutting concerns into aspect.
Cross-cutting concerns are issues that span multiple modules, such as logging, security, and error handling.
In traditional programming techniques like object-oriented programming, cross-cutting concerns are scattered across classes and functions, decreasing code reusability and maintainability. \ac{AOP} defines these cross-cutting concerns as \textit{aspects} separate from the main business logic.
One of the primary models in \ac{AOP} is the \ac{JPM}, comprising \ac{JP}, Pointcut, and Advice. A \ac{JP} refers to a specific point in the program, such as method calls, method executions, field accesses, and exception handling.
A Pointcut specifies the conditions for creating a \ac{JP}. Advice defines the code executed at the \ac{JP} matched by the Pointcut.
\ac{JPM} is used in implementations such as AspectJ and AspectC++.

\subsection{What is Julia}

Julia is a high-level, high-performance dynamic programming language developed by Jeff Bezanson et al.~\cite{bezanson2012juliafastdynamiclanguage, doi:10.1137/141000671}.
While it is a scripting language, it incorporates \ac{JIT} compilation and caching during execution, achieving the performance of compiled languages.
Julia also has functional programming features, supporting higher-order functions, closures, and anonymous functions.
Additionally, it has a convenient \ac{AST} manipulation system and macro features.
Furthermore, Julia offers multiple dispatch, a dynamic type system, and a rich ecosystem of packages, making it widely used for numerical and scientific computing.

\section{AspectJulia: Integration of \ac{AOP} into Julia}

Julia's primary external method of changing program behavior is using macros with metaprogramming.
However, macros directly manipulate Julia's \ac{AST}, so it is necessary to understand the program syntax and have the skills to manipulate it, even though there are packages like MacroTools~\cite{FluxMLMa12:online} that simplify operations for specific syntax.
Integrating \ac{AOP} into Julia makes it possible to rewrite program behavior with a shallow understanding of the syntax using the \ac{AOP} concept.
We have developed a package called \textit{AspectJulia} to achieve this.

One disadvantage of general \ac{AOP} languages is the difficulty of editing code without fully understanding the original code and aspects.
AspectJulia mitigates it by providing the \ac{AST} of \ac{JP}, allowing more straightforward modifications with basic \ac{JP} knowledge in some cases.
Since AspectJulia is implemented as a regular Julia package, users can immediately benefit from new features in Julia without waiting for updates, unlike tools like AspectC++.
Moreover, errors in aspects of AspectJulia can be easily debugged by a stack trace of Julia's native function.
Another advantage of Julia as the base for \ac{AOP} implementation is that Julia's \ac{JIT} optimization can reduce the performance overhead caused by \ac{AOP}-related code.
\textbf{
  While many other aspect-oriented languages restrict \ac{JP}s to specific actions such as method calls or field access, AspectJulia allows a broader range of syntax to be specified as \ac{JP}s, including code blocks and expressions.
  This flexibility is further enhanced by Julia's ability to receive the specified code as an \ac{AST} rather than merely as objects or pointers.
  In Julia, as in many Lisp-based languages, code is treated as data, allowing easy code manipulation within a program without external tools.
  It represents a clear advantage compared to \ac{AOP} extensions based on languages lacking such functionality.}
It is typically challenging for existing aspect-oriented languages to treat operations, such as generating \ac{JP}s for nested loops that meet specific conditions and swapping the matched loops.
For AspectJulia, however, it is feasible.

\section{Related work}

\subsection{Context-oriented Programming}

\ac{COP} is a programming paradigm proposed by Roger Keays et al.~\cite{10.1145/940923.940926}.
\ac{COP} is sometimes described as an extension of \ac{AOP}.
While \ac{AOP} is often considered to encompass the functionalities of \ac{COP}, \ac{COP} introduces a distinct approach to handling cross-cutting concerns~\cite{JOT:issue_2008_03:article4}.
The core idea of \ac{COP} is to provide a way to adapt software's behavior based on its execution context dynamically.
Unlike AOP, which separates cross-cutting concerns using \ac{JP} and Advice, \ac{COP} introduces \textit{Layer}s.
These \textit{Layer}s can be activated or deactivated to manage cross-cutting concerns dynamically.
There are implementations of \ac{COP}, such as ContextJ~\cite{ContextJ}.
However, \ac{COP} has challenges, such as performance issues and difficulties in debugging due to its dynamic context evaluation.

\subsection{Spring Framework}

Spring Framework~\cite{SpringFr13:online} is an open-source application framework by Rod Johnson et al. that provides a comprehensive programming and configuration model for Java applications.
It is widely used in web development.
It includes an \ac{AOP} framework called Spring AOP, which enables developers to define \textit{aspects} using either annotation-based configurations, similar to AspectJ, or schema-based definitions with XML (as shown in Code~\ref{listing:springaop}).

\begin{code}
  \begin{minted}{xml}
<aop:config>
  <aop:aspect id="aspect1" ref="aBean">
    <aop:pointcut id="pointcut1"
      expression="execution(* com.xyz.*.*(..))"/>
    <aop:before pointcut-ref="pointcut1" 
      method="monitor"/>
    ...
  </aop:aspect>
</aop:config>
\end{minted}
  \caption{Example of schema-based definition in Spring AOP.}
  \label{listing:springaop}
\end{code}

\subsection{AspectJ}

AspectJ is an aspect-oriented extension to Java developed by Gregor Kiczales et al.~\cite{10.1007/3-540-45337-7_18}.
It provides a way to modularize cross-cutting concerns by defining \textit{aspects} that can alter the behavior of Java classes.
AspectJ allows for source-level weaving and bytecode-level weaving.
AspectJ is utilized to implement certain features within Spring AOP.
Bruno Harbulot et al.'s research~\cite{10.1145/976270.976286} provides an example of using AspectJ in HPC applications.

\subsection{AspectC++}

AspectC++ is an aspect-oriented extension to C++ developed by Olaf Spinczyk et al.~\cite{10.5555/564092.564100}.
%Introducing concepts similar to AspectJ enables the separation of cross-cutting concerns.
AspectC++ code is converted to native C++ code using AspectC++'s transcompiler.
Robert Clucas et al.'s research~\cite{10.1145/2815782.2815818} and Osamu Ishimura et al.'s~\cite{9835203} provide examples of using AspectC++ in HPC applications.

\subsection{ACC}

ACC is an aspect-oriented extension to C developed by Weigang Gong et al.~\cite{acc}.
It is implemented as a transpiler for the C language.
As of 2024, development has been discontinued and is no longer publicly available.

\subsection{Aclang}

Aclang is a specification for the aspect-oriented extension to C developed by Zhe Chen et al.~\cite{10.1145/3649834}.
It is implemented as part of the analysis tool, movec~\cite{drzchenm59:online}.
It is unrelated to the same named AClang LLVM compiler for OpenMP~\cite{AClang–A62:online}.

\subsection{Cassette}

Cassette~\cite{JuliaLab27:online} is a Julia package that provides functionality similar to the context-based dispatching framework.
Jarrett Revels et al. initially developed it under Capstan~\cite{JuliaDif95:online}, designed for automatic differentiation~\cite{revels2018dynamic}.
Cassette allows developers to define custom execution contexts by modifying CodeInfo, an object in Julia that holds the intermediate representation.

\section{Design overview}

We implemented AspectJulia as a \ac{JPM} acting on the \ac{AST} statically.
Namely, it directly recognizes (Pointcut and \ac{JP}s) and modifies (Advice) the static source code in \ac{AST}.
AspectJulia does not capture dynamically generated functions or directly consider runtime scopes. (We may program Advice to do these when invoked.)
% This design works well when a concern in the source code is expressed with consistent symbol names for objects, a natural and sound convention in programming.

In the design of AspectJulia, overhead in the transformation and execution are given minimal consideration and are not a primary focus.
HPC applications, AspectJulia's primary target, typically have much longer execution times than the time spent on compilation and initialization.
Furthermore, the runtime performance is expected to be close to the pure overhead of the weaved logic, as Julia itself optimizes performance through \ac{JIT} compilation and caching during execution.

\section{Implementation}

\subsection{Pointcut and \ac{JP}}

\subsubsection{Standard Pointcut definition}

In AspectJulia, there are two types of Pointcuts:

\begin{itemize}
  \item Internal Pointcut: Modifies the object's behavior, such as the behavior of functions or the definition of modules.
  \item External Pointcut: Alters the invocation of objects, such as function calls or field accesses.
\end{itemize}

A \ac{JP} passed to Advice varies depending on its originating Pointcut.
However, it typically contains data such as the original expression, the types of the arguments, and information about what Pointcut generated it.

As the implemented Internal Pointcuts, the following are listed with the explanations and their corresponding JPs to be passed to Advices.

\begin{itemize}[first=\footnotesize]
  \item @PCExecFunc:
        \textit{Function definition.}
        $\to$ JPExecFunc
  \item @PCModule:
        \textit{Module definition.}
        $\to$ JPModule
  \item @PCStruct:
        \textit{Structure definition.}
        $\to$ JPStruct
\end{itemize}

Implemented External Pointcuts are the following:

\begin{itemize}[first=\footnotesize]
  \item @PCCallFunc:
        \textit{Function calls.}
        $\to$ JPCallFunc
  \item @PCAssign:
        \textit{Variable assignments.}
        $\to$ JPAssign
  \item @PCAssignAry:
        \textit{Array element assignments.}
        $\to$ JPAssign
  \item @PCAssignSt:
        \textit{Field assignments.}
        $\to$ JPAssign
  \item @PCRefAry:
        \textit{Array element references.}
        $\to$ JPRef
  \item @PCRefSt:
        \textit{Field references.}
        $\to$ JPRef
\end{itemize}

The difference between an external Pointcut and an internal Pointcut is where the \ac{JP} is generated.
This distinction leads to significant differences in their semantics.
An internal Pointcut captures an object's definition (behavior).
Therefore, when the object is duplicated with another symbol, the duplicated with another symbol will have the exact modified definition by an Advice associated with the Pointcut.
On the other hand, an external Pointcut captures the invocation of an object with a specified symbol.
Therefore, if the object's symbol changes, the modification by an external Pointcut is no longer applied.

This is illustrated by the examples in Code~\ref{listing:internal} and Code~\ref{listing:external}.
Consider the difference between @PCExecFunc and @PCCallFunc for \mintinline{julia}{:foo}.
In the Advice, both insert \mintinline{julia}{print("b!")}.
In the case of @PCExecFunc, since it alters the behavior of the \mintinline{julia}{foo} function itself, the Advice is still applied and executed even if the function is defined with a different name; conversely, in @PCCallFunc, which changes how the \mintinline{julia}{foo} function is called, renaming the function results in the Advice not being applied.

\begin{figure}
  \begin{minipage}{0.49\columnwidth}
    \centering
    \begin{minted}{julia}
function foo() 
  # JP here
  print("b!")
  println("foo")
end

function main()
  bar = foo
  foo()
  bar()
end

main() #> b!foo\nb!foo
\end{minted}
    \captionof{code}{Internal Pointcut.}
    \label{listing:internal}
  \end{minipage}
  \begin{minipage}{0.49\columnwidth}
    \centering
    \begin{minted}{julia}
function foo()
  println("foo")
end

function main()
  bar = foo
  # JP here
  print("b!")
  foo() 
  bar()
end

main() #> b!foo\nfoo
\end{minted}
    \captionof{code}{External Pointcut.}
    \label{listing:external}
  \end{minipage}
\end{figure}

In addition to the above, there is @PCAttr.
With it, \ac{JP}s are generated at the specified position on the source code using the \mintinline{julia}{@attr} keyword.
The types of \ac{JP} generated are automatically determined based on the specified \ac{AST} nodes.

\subsubsection{Pointcut Matching for Standard Pointcut}

All Pointcuts take a pattern as an argument. The pattern can be specified using a Symbol~(for exact match) or String~(for partial match).
For the simplicity of the implementation, the current matching is performed only on individual elements.
Namely, it does not support complex conditions, like destructuring assignments and composite types.

\subsubsection{Argument type matching for Standard Pointcut}

In AspectJulia, extracting information about arguments as a named tuple and performing matching against the definitions in the Pointcut is possible.
It supports variable names, keyword arguments, variable types, and variadic arguments.
Currently, it is only supported by @PCExecFunc.
It does not support composite types.

\begin{itemize}
  \item \mintinline{julia}{Axxx([symbol])}: It represents an argument of type ``xxx''. If ``xxx'' is ``Any'', it can accept any type. When a symbol is specified, it matches the specified symbol. When only the type name is given (\mintinline{julia}{Axxx}), it is equivalent to (\mintinline{julia}{Axxx()}).
  \item \mintinline{julia}{VAxxx([symbol])}: It represents a variable-length argument of type ``xxx''. If ``xxx'' is ``Any'', it can accept any type. If a symbol is specified, it matches the symbol. When only the type name is given (\mintinline{julia}{VAxxx}), it is equivalent to \mintinline{julia}{VAxxx()}.
  \item \mintinline{julia}{KAxxx(symbol)}: It represents a keyword argument of type ``xxx''. If ``xxx'' is ``Any'', it can accept any type. The symbol is required.
  \item \mintinline{julia}{KVAAny(symbol)}: It represents variable-length keyword arguments.
\end{itemize}

Code~\ref{listing:arg} shows an example of using this feature.

\begin{code}
  \begin{minted}{julia}
function foo(a::Int64) end
@PCExecFunc(:foo, [AAny]) # Matched
@PCExecFunc(:foo, [AInt64]) # Matched
@PCExecFunc(:foo, [AFloat64]) # Not matched
@PCExecFunc(:foo, [AInt64(:b)]) # Matched
@PCExecFunc(:foo, [AInt64(:a)]) # Not matched

function bar(a, as::Int64...; z::Int64) end
@PCExecFunc(:bar, [AAny, VAInt64, KAAny(:z)] ) # Matched
\end{minted}
  \caption{Argument type matchings.}
  \label{listing:arg}
\end{code}

\subsubsection{XML-like Pointcut Definition}
To specify complex conditions for a Pointcut targeting specific functions, PCXPath is implemented, which allows for the specification of paths used in internal matching with xpath~\cite{Coverpag10:online}-like syntax.

The following is a list of nodes and their attributes.
The brackets ``[]'' denote optional attributes.

\begin{itemize}
  \item joinpoint: Root node
  \item macro: Macro declaration / @name, [@attr]
  \item macrocall: Macro call / @name, [@attr]
  \item module: Module declaration / @name, [@bare], [@attr]
  \item function: Function declaration / @name, [@args], [@attr]
  \item call: Function call / @name, [@ref], @argc, [@parallel], [@attr]
  \item struct: Struct declaration / @name, [@mutable], [@attr]
  \item ref: Reference / @name, [@ref], [@attr]
  \item assign: Assignment / @name, [@ref], [@op], [@attr]
  \item for: for loop / @iterc, [@comprehension], [@attr]
\end{itemize}

The ref attribute represents an index for references, denoted with ``\%'' in place of the name:

\begin{itemize}
  \item \mintinline{julia}{[]}: Array access
  \item \mintinline{julia}{.x}: Field access (where \textit{x} is the field name)
  \item \mintinline{julia}{(,,)}: Tuple access
\end{itemize}

References can be expressed in combination with these notations.
For example,``\mintinline{julia}{(,,%[].x[])}'' appears as a ref in \mintinline{julia}{(_, _, ary[1].x[2]) = val}.

\subsection{Advice}

In AspectJulia, there are two categories of Advice:
\begin{itemize}
  \item Insert Advice: This type of Advice inserts code at the \ac{JP}. The original logic remains unmodified. It means that Insert Advice does not modify arguments or alter return values of the original code.
  \item Replace Advice: This type of Advice partially or entirely replaces the \ac{JP}. \textbf{A key feature of AspectJulia Replace Advice is that it can receive and manipulate the \ac{JP}'s \ac{AST}. It allows for direct and flexible modifications to the original code.}
  \item
\end{itemize}

In contrast, in AspectJ and AspectC++, Advice only can execute the \ac{JP}'s logic through a given object representing the \ac{JP}, limiting the ability to analyze or modify the logic.

\subsubsection{Insert Advice Types}

Insert Advice receives a \ac{JP} and returns the \ac{AST} to insert.
The following types are categorized into Insert Advice.
The Advice that ends with ``A'' receives the \ac{JP}'s arguments.

\begin{itemize}[first=\footnotesize]
  \item @ADBefore / @ADBeforeA:
        \textit{Executes before the \ac{JP}.}
  \item @ADAfterR / @ADAfterRA:
        \textit{Executes after the \ac{JP}, receiving \ac{JP}'s return value.}
  \item @ADAfterThrowing / @ADAfterThrowingA:
        \textit{Executes if the \ac{JP} throws an exception, receiving the exception.}
  \item  @ADAfter / @ADAfterA:
        \textit{Executes after the \ac{JP}, regardless of whether an exception occurred.}
\end{itemize}

\subsubsection{Replace Advice Types}

Replace Advice receives a \ac{JP} and returns the \ac{AST} to replace by.
The following types are categorized into Replace Advice.

\begin{itemize}[first=\footnotesize]
  \item @ADAround:
        \textit{Replaces the \ac{JP}.}
  \item  @ADAppendF / @ADAppendB:
        \textit{Inserts an \ac{AST} before/after the \ac{JP}.}
\end{itemize}

\subsubsection{Another Advice Type}

\begin{itemize}[first=\footnotesize]
  \item @ADNothing:
        \textit{Specifies empty Advice.}
\end{itemize}

\subsubsection{Fusion of Advice}

The syntax in Code~\ref{listing:fusion} can be used to compose Advice.

\begin{figure}
  \begin{minipage}{0.49\columnwidth}
    \centering
    \begin{minted}{julia}
@ADBefore() & @ADAfter()


\end{minted}
    \captionof{code}{Fusion of Advice.}
    \label{listing:fusion}
  \end{minipage}
  \begin{minipage}{0.49\columnwidth}
    \centering
    \begin{minted}{julia}
Expr(
:aj, JoinPointParam(...), 
e.head, e.args...)
\end{minted}
    \captionof{code}{aj Node declaration.}
    \label{listing:aj}
  \end{minipage}
\end{figure}

\subsection{Aspect}

The design of \textit{aspects} in AspectJulia differs from languages like AspectC++ and AspectJ in two points.

First, while AspectC++ and AspectJ allow an \textit{aspect} to include multiple Pointcuts and Advice, an \textit{aspect} in AspectJulia comprises a single Pointcut and a single Advice.
This restriction in AspectJulia is intended to simplify the resolution of order when multiple \textit{aspects} are used simultaneously.
In practice, AspectC++ faces challenges related to \textit{aspect} inheritance (analogous to class inheritance) and the definition of order when determining the application sequence of \textit{aspects} and \textit{advice} when multiple \textit{aspects} are involved.
These complexities can lead to issues in ensuring the correct application sequence.

Second, unlike AspectC++ and AspectJ, an \textit{aspect} of AspectJulia cannot hold its own resources because Julia, unlike C++ or Java, is not an object-oriented language.
Therefore, any necessary resources must be added to global variables or the specified objects, with the user responsible for managing them.
This approach reduces the internal complexity of \textit{aspects} implementation but requires careful resource management by the developer.

\subsection{Weave and Emit}

The code transformation process in AspectJulia has three stages: pre-weave, weave, and emit. The details are described in the following sections.
In the following sample codes, the \mintinline{julia}{f_adv_xx} function represents a function generated by the Advice, and \mintinline{julia}{original_expr} represents the specified code's original \ac{AST}.

\subsubsection{[Pre-weave] Include functions resolve}

In Julia, the \mintinline{julia}{include} function dynamically incorporates code from other files.
However, AspectJulia supports rewriting only static code, so it cannot support dynamically added code during the execution.
An exception is made to \mintinline{julia}{include} calls with a String argument, which are processed before weaving.
This exception accommodates Julia's standard practice of splitting a module across multiple files and using \mintinline{julia}{include} to consolidate them within the module declaration.

\subsubsection{[Pre-weave] Expand Attribution Macro for AspectJulia}

When expanding the \mintinline{julia}{@attr} keyword for @PCAttr, the macro appends an AttributeNode to the args of the specified \mintinline{julia}{Expr} on the \ac{AST}.

\subsubsection{[Weave] Generate \ac{JP}s}

In the weaving stage, \ac{JP}s are generated using a crawler derived from a Pointcut, traversing the \ac{AST}.
When node \mintinline{julia}{e} is identified as a \ac{JP}, it is replaced with a node shown in Code~\ref{listing:aj} (called aj Node).
\mintinline{julia}{JoinPointParam} is information about the \ac{JP}.

\subsubsection{[Emit] Convert \ac{AST} with aj Node to Standard \ac{AST}}

An \ac{AST} that contains aj Nodes, which indicate a \ac{JP}, cannot be executed as-is and must be transformed into a standard \ac{AST}.
This transformation is performed by considering the specific Advice that will be applied to the \ac{JP}.
The transformation rules are detailed in the following sections.

\subsubsection{[Emit] Receiving a return value of a statement}

An @ADAfterRunning receives the return value from the \ac{JP}. Consequently, the syntax is expanded, as Code~\ref{listing:2}.

\subsubsection{[Emit] Pre-evaluation of argument variables}

When dealing with \ac{JP}s arguments, it is necessary to pre-evaluate variables that the specified expression receives.
However, if the evaluation of these variables has side effects, there is a risk that the introduction of Advice could alter the value of the arguments passed to the function.

For example, consider the code in Code~\ref{listing:3}.

\begin{figure}
  \begin{minipage}{0.49\columnwidth}
    \centering
    \begin{minted}{julia}
let result = original_expr
  f_adv(result)
  result
end
\end{minted}
    \captionof{code}{Receiving a return value.}
    \label{listing:2}
  \end{minipage}
  \begin{minipage}{0.49\columnwidth}
    \centering
    \begin{minted}{julia}
ary = [1, 2, 3]
foo(pop!(ary)) # foo(3)


\end{minted}
    \captionof{code}{Function call with an argument with side effects.}
    \label{listing:3}
  \end{minipage}
  \begin{minipage}{0.49\columnwidth}
    \centering
    \begin{minted}{julia}
ary = [1, 2, 3]
f_adv(pop!(ary)) # f_adv(3)
foo(pop!(ary)) # foo(2)


\end{minted}
    \captionof{code}{Insert Advice to Code~\ref{listing:3}.}
    \label{listing:4}
  \end{minipage}
  \begin{minipage}{0.49\columnwidth}
    \centering
    \begin{minted}{julia}
ary = [1, 2, 3]
let arg1 = pop!(ary)
  f_adv(arg1)
  foo(arg1)
end
\end{minted}
    \captionof{code}{Pre-evaluation of argument variables.}
    \label{listing:5}
  \end{minipage}

\end{figure}

If we introduce an \textit{advice} as Code~\ref{listing:4}, the value passed to the function will differ from the original code.

Code~\ref{listing:4} demonstrates that the \textit{advice} can alter the value passed to the \mintinline{julia}{foo} function, which might result in unintended behavior.
Code~\ref{listing:5} shows that the value should be pre-evaluated and stored in a variable before applying the \textit{advice} to prevent unintended behavior.

However, this pre-evaluation of argument variables may lead to unintended behavior when the specified function is invoked with the short-circuit logic.
For example, when a \ac{JP} is created at the \mintinline{julia}{@attr} ``p'' attribute on Code~\ref{listing:shortcircuit}, which contains short-circuit logic and performs a pre-evaluation, the code is transformed into Code~\ref{listing:shortcircuit2}.
This modification reveals that the resulting behavior of the code has changed.
Because this case is not commonly encountered in typical Julia codes, the present implementation of AspectJulia leaves the handling of such cases to users.

\begin{code}
  \begin{minted}{julia}
ary = [false, true, false]
@attr "p" pop!(ary) && pop!(ary)
println(ary) #> Bool[0, 1] ([false, true])
\end{minted}
  \caption{Sample Code with short-circuit logic.}
  \label{listing:shortcircuit}
\end{code}

\begin{code}
  \begin{minted}{julia}
ary = [false, true, false]
let arg1 = pop!(ary), arg2 = pop!(ary)
  result = arg1 && arg2
  result
end
println(ary) #> Bool[0] ([false])
\end{minted}
  \caption{Add pre-evaluation to Code~\ref{listing:shortcircuit}.}
  \label{listing:shortcircuit2}
\end{code}

When the specified expression involves variable assignment, it is possible that the expression also serves as the declaration of that variable.
Since variables declared within a let block are not registered in the external scope, checking whether the variable already exists is necessary.
The variable should be declared before the let block if it does not exist. Therefore, the code would be expanded as Code~\ref{listing:6}.

\begin{code}
  \begin{minted}{julia}
if !@isdefined(var_x) # @isdefined is a built-in Julia 
  var_x = nothing     # macro for variable existence
end                   # check.
let arg1 = f()
  f_adv(arg1)
  var_x = arg1
end
\end{minted}
  \caption{Existence check of a variable by @isdefined macro.}
  \label{listing:6}
\end{code}

Variable assignments that include checks for the variable's existence can introduce significant overhead, making this approach generally inadvisable.
In Julia, the assignment operator cannot be overloaded; therefore it always returns the assigned value.
It means pre-evaluating arguments are typically unnecessary.
As a result, using AfterRunning, which captures only the return value, is usually sufficient for handling most cases without the need for complex pre-evaluation logic.

Due to limitations in the current implementation of static analysis, AspectJulia does not support the use of global variables.
Therefore, global variables are not considered in this step.

\subsubsection{[Emit] Add exception handling}

When inserting @ADAfter and @ADAfterThrowing, it is necessary to insert a try-catch-finally structure.
The expanded syntax would look like Code~\ref{listing:7}.

\begin{code}
  \begin{minted}{julia}
try
  original_expr
catch e
  f_adv_throwing(e)
  throw(e)
finally
  f_adv_after()
end
\end{minted}
  \caption{Exception handling}
  \label{listing:7}
\end{code}

Because @ADAfter and @ADAfterThrowing are implemented as Insert Advice, the exception is not suppressed, maintaining the integrity of the program's original exception-handling logic.
%However, throwing a different exception within the Advice is possible.

\subsection{Using AspectJulia}

At first, load the AspectJulia package.
Next, call the \mintinline{julia}{setupasp} function.
The \mintinline{julia}{setupasp} function takes an array of Aspect as arguments.
%Additionally, it is possible to specify a config of type AspConfig and the \mintinline{julia}{scope::Module} to be registered.
The definition and behavior can be modified using the config for the \mintinline{julia}{setupasp}.
Modifying the \mintinline{julia}{toolsetname_header} in the config changes the names of macros and functions.
Code~\ref{listing:sample} shows a sample code to initialize AspectJulia.

This code loads the AspectJulia module and calls the \mintinline{julia}{setupasp} function.
The \mintinline{julia}{setupasp} registers an \textit{advice} to output the string "before foo" before calling the \mintinline{julia}{:foo} function.
As a result, the following function is registered in the environment.

\begin{itemize}
  \item \mintinline{julia}{@asp}: Weave to the given code and execute it.
  \item \mintinline{julia}{asp_src}: Weave the code of the given path and execute it.
  \item \mintinline{julia}{asp_code}: Weave to the given \ac{AST} and execute it.
  \item \mintinline{julia}{@asp_nr}: Weave to the given code and return the modified \ac{AST}.
  \item \mintinline{julia}{asp_src_nr}: Weave the code of the given path and return the modified \ac{AST}.
  \item \mintinline{julia}{asp_code_nr}: Weave to the given \ac{AST} and return the modified \ac{AST}.
\end{itemize}

% The AspConfig type has the following fields:

% \begin{itemize}
%   \item \mintinline{julia}{preserve_module}:
%         Whether to register modules not woven into the environment. If true, one of the \mintinline{julia}{newname_prefix}, \mintinline{julia}{newname_suffix} fields must be specified.
%   \item \mintinline{julia}{newname_prefix}:
%         Adds the provided \mintinline{julia}{Symbol} as a prefix to the name of the woven module.
%   \item \mintinline{julia}{newname_suffix}:
%         Adds the provided \mintinline{julia}{Symbol} as a suffix to the name of the woven module.
%   \item \mintinline{julia}{preserve_linenumbernodes}:
%         In Julia, the source code information is included in the \ac{AST} as \mintinline{julia}{LineNumberNode}.
%         If true, the \mintinline{julia}{LineNumberNode} present in the \ac{AST} being woven is preserved. (Note: \mintinline{julia}{LineNumberNode}s present in Pointcut and Advice are always retained.)
%   \item \mintinline{julia}{toolsetname_header}:
%         The seed for the names of macros and functions.
%         The default is \mintinline{julia}{:asp}.
%   \item \mintinline{julia}{debug_printgeneratedsyntaxtree}:
%         Whether to output the syntax tree registered by the AspectJulia system.
%         If \mintinline{julia}{preserve_module} is true, the output syntax tree will include both the pre-weave and post-weave versions.
% \end{itemize}

% The \mintinline{julia}{setupasp} function registers the following objects in the scope when using the default configuration.

\begin{code}
  \begin{minted}{julia}
using .AspectJulia

setupasp([Aspect(
  @PCCallFunc(:foo), 
  @ADBefore((tjp) -> :(() -> print("before foo!"))))])

@asp module Test
  function foo()
    print("foo")
  end
  function main()
    foo()
  end
end

Test.main() #> before foo!foo
\end{minted}
  \caption{Sample code to set up AspectJulia.}
  \label{listing:sample}
\end{code}

\subsection{Debug capabilities}

Many existing \ac{AOP} extension languages need help in debugging.
AspectJulia offers features to mitigate these challenges.
% When using the \mintinline{julia}{debug_printgeneratedsyntaxtree} option, the code generated after weaving includes debugging information utilizing \mintinline{julia}{LineNumberNode}.
In the \mintinline{julia}{LineNumberNode}, additional information is embedded in the filename section, indicating which Pointcut generated the \ac{JP}.
Moreover, if the matching of @PCExecFunc fails solely due to differences in type specifications of arguments, debug information will display the relevant Pointcut and the location of the original code.
This helps identify where and why the matching did not occur.
When applying the aspect defined in Code~\ref{listing:sample} saved as ``Sample.jl'' to ``Test.jl'', the \ac{AST} registered in the environment will be shown in Code~\ref{listing:debug}.

\begin{code}
    \begin{minted}{julia}
module Test
#= Test.jl:10 =#
#= Test.jl:11 =#
function foo()
    #= Test.jl:11 =#
    #= Test.jl:12 =#
    println("foo")
end
#= Test.jl:14 =#
function main()
    #= Test.jl:14 =#
    #= Test.jl:15 =#
    begin
        #= AOP: PCCallFunc(:foo) ##= Sample.jl:2 =##:0 =#
        ((()->begin
                #= Sample.jl:5 =#
                println("before call")
            end))()
        begin
            foo()
        end
    end
end
end
\end{minted}
    \caption{Debug information.}
    \label{listing:debug}
\end{code}

\subsection{Advice on Advice}

Advice on Advice is not directly supported.
However, it can be achieved by recursively applying the modification (weave $\to$ emit) processes.
Code~\ref{listing:adviceonadvice} shows an example of this.

\begin{code}
  \begin{minted}{julia}
# The first aspect is registered as macros 
# and functions with the suffix 'asp1'.
setupasp([];                                     
 config = AspConfig(toolsetname_header = :asp1)) 

# The second aspect is registered as macros
# and functions with the suffix 'asp2'.
setupasp([];                                      
 config = AspConfig(toolsetname_header = :asp2))

AspectJulia.parse(path) |> 
asp1_code_nr |> # xx_code_nr is a function that receives
asp2_code_nr |> # an AST and returns a modified.
AspectJulia.register # Register the modified AST.
\end{minted}
  \caption{Advice on Advice.}
  \label{listing:adviceonadvice}
\end{code}

\subsection{Limitations and unimplemented features}

In some cases, Julia offers different syntaxes to define the same behavior (for instance, defining functions; \mintinline{julia}{function f() ... end}, \mintinline{julia}{f() = ...} and \mintinline{julia}{f = () -> ...}).

AspectJulia, which analyzes the \ac{AST} directly, treats these different ways as distinct, even though they may perform the same function.
To consistently handle these variations, we would need to create specific rules (Pointcuts) for each variation or enhance the crawlers' recognition of them.
However, the current version of AspectJulia only covers a few of these variations and does not provide an easy way to define such Pointcuts.

One reason is that the current implementation of AspectJulia is still experimental.
Another reason is the design concept of AspectJulia:
\textbf{AspectJulia does not just focus on the names of functions or objects; it also considers how they are written.
  For instance, programmers often use shorthand for temporary or local functions and more formal definitions for long-lasting or global ones.
  AspectJulia is designed to recognize and account for these differences in expression.}

\section{Feature comparison}

Table~\ref{tab:comparison} compares AspectJulia with Cassette, AspectJ and AspectC++ in terms of key features.
Spring AOP is not included in the comparison because it is not a language extension.

\begin{table*}[htbp]
  \centering
  \caption{Feature comparison of AspectJulia, Cassette, AspectJ, and AspectC++.}
  \label{tab:comparison}
  \begin{tabular}{l|l|l|l|l}
                                                          & AspectJulia                 & Cassette       & AspectJ                & AspectC++       \\ \hline\hline
    Based language                                        & Julia                       & Julia          & Java                   & C++             \\
    Model                                                 & AOP/JPM                     & Similar to COP & AOP/JPM                & AOP/JPM         \\ \hline
    Dynamic matching                                      & no                          & yes            & yes (cflow)            & yes (cflow)     \\
    Scope handling                                        & yes                         & yes (context)  & yes                    & yes             \\
    Function implementation modification                  & yes                         & yes (context)  & yes                    & yes             \\
    Function call modification                            & yes                         & yes (context)  & yes                    & yes             \\
    Intervention in a variable assignment                 & yes                         & Not documented & yes                    & yes             \\
    Intervention in a variable reference                  & partial (Only array/struct) & Not documented & yes                    & yes             \\
    Intervention in other code blocks                     & yes                         & Not documented & external (e.g. LoopAJ) & no              \\ \hline
    Modification of AST at \ac{JP}                        & yes                         & yes            & no                     & no              \\ \hline
    Preservation of states of the extension object itself & no                          & yes            & yes                    & yes             \\
    Order of extension objects                            & yes                         & -              & yes                    & yes             \\
    Inheritance of extension object                       & no                          & -              & yes                    & partial (Slice)
  \end{tabular}
\end{table*}

\section{Use cases}

We will demonstrate the minimal conceptual code for each use case.
Each use case presents the module implementing the primary concern, the Aspect, and the code after weaving.
For the sake of simplicity, the initialization, \mintinline{julia}{setupasp}, the \mintinline{julia}{LineNumberNode}, module declarations, and redundant \mintinline{julia}{:block} are omitted from the example codes.

\subsection{Logging, dump, and assertion}
In aspect-oriented languages, a typical example of a cross-cutting concern is the separation of logging.
It can be implemented in AspectJulia using Insert Advice.

Insert Advice can also be applied to dump the arguments and results of function calls or to add argument checks and perform assertions, which is particularly useful in business applications such as access control.
%Replace Advice should be used when arguments need to be modified, which will be discussed separately.

In the use case, we will add cross-cutting concerns such as call logging, return value logging, error logging, and zero-division checks to the \mintinline{julia}{mycalc} function shown in Code~\ref{listing:logging_mod} and Code~\ref{listing:logging}.
In the scenario, call logging and argument checks will be performed using @ADBeforeA, return value logging will be handled by @ADAfterR, and exception logging will be managed by @ADAfterThrowing.

\begin{code}
  \begin{minted}{julia}
function mycalc(x, y, z=100)
  (x + y) / z 
end
\end{minted}
  \caption{Target for the use case for logging, dump, and assertion.}
  \label{listing:logging_mod}
\end{code}

\begin{code}
  \begin{minted}{julia}
Aspect(
  @PCExecFunc(:mycalc),
  @ADBeforeA((tjp::JPExecFunc) -> :((arg) ->begin
    println("exec $($(tjp.name)) with $arg")
    if arg.args[3] == 0
      error("zero division")
    end
  end)) & @ADAfterR((tjp::JPExecFunc) -> :((result) -> 
println("$($(tjp.name)) return $result"))) &
  @ADAfterThrowing((tjp::JPExecFunc) ->
   :((exception) -> println("exception $exception"))))
\end{minted}
  \caption{Aspect for logging, dump, and assertion.}
  \label{listing:logging}
\end{code}

\begin{code}
  \begin{minted}{julia}
function mycalc(x, y, z = 100)
  try
    ((arg->begin
        println("exec $(mycalc) with $(arg)")
        if arg.args[3] == 0
          error("zero division")
        end
        end))((args = [x, y, z], kargs = Dict([])))
    resulttmp = (x + y) / z
    ((result->
println("$(mycalc) return $(result)")))(resulttmp)
    resulttmp
  catch e
    ((exception->println("exception $(exception)")))(e)
    throw(e)
  end
end
\end{minted}
  \caption{Generated code from Code~\ref{listing:logging_mod} and Code~\ref{listing:logging}.}
  \label{listing:logging_weaved}
\end{code}

\subsection{Performance Profiling}

In the use case, we implement an \textit{aspect} to measure the execution time of the \mintinline{julia}{bar} function shown in Code~\ref{listing:profiling} by using Replace Advice.
Specifically, @ADAround wraps the function call with the \mintinline{julia}{@time} macro.
This approach can also be applied to insert other debugging or optimization macros.

\begin{code}
  \begin{minted}{julia}
bar() = sleep(10)
function main()
  bar()
end
\end{minted}
  \caption{Target for the use case for profiling.}
  \label{listing:profiling_mod}
\end{code}

\begin{code}
  \begin{minted}{julia}
Aspect(
  @PCCallFunc(:bar),
  @ADAround((tjp::JPCallFunc, original_expr) -> quote
    @time $original_expr
  end))
\end{minted}
  \caption{Aspect for profiling.}
  \label{listing:profiling}
\end{code}

\begin{code}
  \begin{minted}{julia}
bar() = sleep(10)
function main()
    @time bar()
end
\end{minted}
  \caption{Generated code from Code~\ref{listing:profiling_mod} and Code~\ref{listing:profiling}.}
  \label{listing:profiling_weaved}
\end{code}

\subsection{Modification of external packages}

It is possible to weave and modify the behavior of external modules.
This approch can use to enhance security or to improve performance in business applications, such as altering network access destinations or replacing function calls with vulnerabilities.
The key advantages is the ability to modify entire modules without fully understanding their source code and the flexibility to turn on or off on a per-module basis.

In the example, a network request's destination in the module \mintinline{julia}{GetResource} in Code~\ref{listing:modify_mod} is changed to ``localhost'' if the target is ``example.net'' using @ADAround in Code~\ref{listing:modify}.

\begin{code}
  \begin{minted}{julia}
module GetResource
function load()
  [myfetch("https://example.net/"),  # myfetch is dummy
   myfetch("https://example.org/"),] # function to fetch
end                                  # network resources
end
\end{minted}
  \caption{Specified external module.}
  \label{listing:modify_mod}
\end{code}

\begin{code}
  \begin{minted}{julia}
Aspect(
  @PCCallFunc(:myfetch),
  @ADAround((tjp::JPCallFunc, original_expr) -> quote
  # extractargs is a utility in AspectJulia for 
  # extracting arguments from AST
  if extractargs(original_expr)[1] ==
                   "https://example.net/"
    myfetch("https://localhost/")
  else
    $original_expr
  end
end))
\end{minted}
  \caption{Aspect to change the behavior of an external module.}
  \label{listing:modify}
\end{code}

\begin{code}
  \begin{minted}{julia}
module GetResource
function load()
  [ if (extractargs(original_expr))[1] ==
        "https://example.net/"
      myfetch("https://localhost/")
    else
      myfetch("https://example.net/")
    end, if (extractargs(original_expr))[1] == 
        "https://example.net/"
      myfetch("https://localhost/")
    else
      myfetch("https://example.org/")
    end]
end
end
\end{minted}
  \caption{Generated code from Code~\ref{listing:modify_mod} and Code~\ref{listing:modify}.}
  \label{listing:modify_weaved}
\end{code}

\subsection{Object extension}

\mintinline{julia}{struct} can be extended to allow custom members to be inserted from external sources.
A similar operation can also be performed on \mintinline{julia}{module}.

In the example, \mintinline{julia}{struct MYST} in Code~\ref{listing:st_mod} is extended to include an \mintinline{julia}{init_time} member and a constructor that sets the initialized time using @ADAppendB in Code~\ref{listing:st}.

\begin{code}
  \begin{minted}{julia}
struct MYST
  x::Int
  y::Int
end
\end{minted}
  \caption{Target for the use case for object extension.}
  \label{listing:st_mod}
\end{code}

\begin{code}
  \begin{minted}{julia}
Aspect(
  @PCStruct(:MYST),
  @ADAppendB((tjp::JPStruct) -> quote
    init_time
    function MYST(x, y)
      new(x, y, mynow()) # mynow is a dummy function 
    end                  # to get the current time
  end))
\end{minted}
  \caption{Aspect for object extension.}
  \label{listing:st}
\end{code}

\begin{code}
  \begin{minted}{julia}
struct MYST
  x::Int
  y::Int
  init_time
  function MYST(x, y)
    new(x, y, mynow()) 
  end                 
end
\end{minted}
  \caption{Generated code from Code~\ref{listing:st_mod} and Code~\ref{listing:st}.}
  \label{listing:st_weaved}
\end{code}

\subsection{Partial weaving using PCXPath}

In cases like recursive functions, it is helpful to apply \textit{advice} only under specific conditions rather than all.
Using PCXPath, \textit{advice} can be applied to function calls that meet particular criteria.
Code~\ref{listing:xpath_target} is an example of a recursive Fibonacci sequence function.
The XML generated from parsing Code~\ref{listing:xpath_target} is shown in Code~\ref{listing:xpath_target_xml}.
In the aspect in Code~\ref{listing:xpath_aspect}, \textit{advice} is weaved only before the call to the \mintinline{julia}{fib} function outside the \mintinline{julia}{fib} function.
The weaved code is shown in Code~\ref{listing:xpath_weaved}.

\begin{code}
  \begin{minted}{julia}
module MyFib
function fib(n)
    if n < 2
        n
    else
        fib(n-1) + fib(n-2)
    end
end
function main()
    println(fib(10))
end
end
\end{minted}
  \caption{Target for the use case for partial weaving.}
  \label{listing:xpath_target}
\end{code}

\begin{code}
  \begin{minted}{xml}
<joinpoint>
  <module name="MyFib">
    <function name="fib" args="n">
      <call name="<" argc="2"/>
      <call name="+" argc="2"/>
      <call name="fib" argc="1"/>
      <call name="-" argc="2"/>
      <call name="fib" argc="1"/>
      <call name="-" argc="2"/>
    </function>
    <function name="main">
      <call name="println" argc="1"/>
      <call name="fib" argc="1"/>
    </function>
  </module>
</joinpoint>
\end{minted}
  \caption{Target for the use case for partial weaving.}
  \label{listing:xpath_target_xml}
\end{code}

\begin{code}
  \begin{minted}{julia}
Aspect(
  @PCXPath(
    "//function[not(contains(@name,'fib'))]" *
    "//call[@name='fib']"
  ),
  @ADBefore((tjp) -> :(() -> println("before call"))))
\end{minted}
  \caption{Aspect for partial weaving.}
  \label{listing:xpath_aspect}
\end{code}

\begin{code}
  \begin{minted}{julia}
module MyFib
function fib(n)
    if n < 2
        n
    else
        fib(n-1) + fib(n-2)
    end
end
function main()
    (() -> println("before call"))()
    println(fib(10))
end
end
\end{minted}
  \caption{Target for the use case for partial weaving.}
  \label{listing:xpath_weaved}
\end{code}

\subsection{Loop swapping}

In AspectJulia, retrieving the \ac{AST} of a \ac{JP} enables advanced manipulations such as reordering loops using Replace Advice.
While prior research, such as LoopAJ~\cite{LoopAJ}, implements a \ac{JP} specifically for loops, AspectJulia offers more flexibility in handling such tasks.
%Although direct control of the \ac{AST} is required, similar techniques can be used to modify function precision, as provided by ChangePrecision packages~\cite{GitHubJu11:online}, using AspectJulia.
%AspectJulia includes several utility functions and, when combined with existing \ac{AST} manipulation packages like MacroTools, can perform even more flexible operations.

In the example shown in Code~\ref{listing:loop_mod} and Code~\ref{listing:loop}, the \mintinline{julia}{myloop} function is modified to swap the order of the loops using @ADAround.
%The \mintinline{julia}{swap_loop} function is a utility that swaps loops in the given \ac{AST}.

Such loop swapping is a common optimization technique in \ac{HPC} applications to improve performance by modifying memory access patterns.

\begin{code}
  \begin{minted}{julia}
function myloop()
  @attr "loopA" for i in 1:10, j in 1:10
    println("x=$i, y=$j")
  end
end
\end{minted}
  \caption{Target for the use case for swapping a loop.}
  \label{listing:loop_mod}
\end{code}

\begin{code}
  \begin{minted}{julia}
Aspect(
  @PCAttr(:loopA),
  @ADAround((tjp::JPDefault, original_expr) ->
  # swap_loop is a utility in AspectJulia for 
  # reversing range objects in the AST
   swap_loop(original_expr))) 
\end{minted}
  \caption{Aspect for swapping a loop.}
  \label{listing:loop}
\end{code}

\begin{code}
  \begin{minted}{julia}
function myloop()
  for j in 1:10, i in 1:10
    println("x=$i, y=$j")
  end
end
\end{minted}
  \caption{Generated code from Code~\ref{listing:loop_mod} and Code~\ref{listing:loop}.}
  \label{listing:loop_weaved}
\end{code}

\section{Conclusion}
This paper presented AspectJulia, a framework integrating \ac{AOP} into Julia.
Julia's powerful metaprogramming and \ac{AST} manipulation capabilities give AspectJulia a broader application scope than other \ac{AOP} extensions.
We demonstrated the usefulness of AspectJulia by providing conceptual implementation examples for basic AOP use cases, a business application scenario, and an HPC application case.
Unlike Cassette, AspectJulia allows additional logic to be inspected directly as Julia code, facilitating learning and simplifying debugging.

Future work will focus on improving the \ac{JP} matching functionality. %and enhancing support for dynamic \mintinline{julia}{include} functions by embedding a crawler.

% \section*{Acknowledgment}

% \section*{Acknowledgment}

% The preferred spelling of the word ``acknowledgment'' in America is without 
% an ``e'' after the ``g''. Avoid the stilted expression ``one of us (R. B. 
% G.) thanks $\ldots$''. Instead, try ``R. B. G. thanks$\ldots$''. Put sponsor 
% acknowledgments in the unnumbered footnote on the first page.

\bibliographystyle{IEEEtran}
\bibliography{mybib}

\end{document}